\documentclass[aps,prb,twocolumn,longbibliography]{revtex4-1}

\usepackage{graphicx}
\usepackage{dcolumn}
\usepackage{hyperref}

\usepackage{bbm}
\usepackage{epsfig}
\usepackage{graphicx}
\usepackage{amsfonts}
\usepackage{amssymb}
\usepackage{amsmath}
\usepackage{bm}
\usepackage{color}
\usepackage{natbib}
\usepackage{overpic}
\usepackage{hyperref}
\hypersetup{colorlinks=true,urlcolor=blue,citecolor=blue}

\begin{document}

\title{Reorthonormalization of Chebyshev matrix product states for dynamical correlation functions}

\author{H. D. Xie$^{1,2}$, R. Z. Huang$^{1,2}$, X. J. Han$^{1,2}$, X. Yan$^1$, H. H. Zhao$^{3}$, Z. Y. Xie$^4$ }

\author{H. J. Liao$^1$}
\email{navyphysics@iphy.ac.cn}

\author{T. Xiang$^{1,2,5}$}
\email{txiang@iphy.ac.cn}

\affiliation{$^1$Institute of Physics, Chinese Academy of Sciences, P.O. Box 603, Beijing
100190, China}

\affiliation{$^2$School of Physical Sciences, University of Chinese Academy of Sciences,
Beijing 100049, China}

\affiliation{$^3$RIKEN Brain Science Institute, Wako-shi, Saitama 351-0198, Japan}

\affiliation{$^4$Department of Physics, Renmin University of China, Beijing 100872, China}

\affiliation{$^5$Collaborative Innovation Center of Quantum Matter, Beijing 100190, China}

\begin{abstract}
  The Chebyshev expansion offers a numerically efficient and easy-implement algorithm for evaluating dynamic correlation functions using matrix product states (MPS).
  In this approach, each recursively generated Chebyshev vector is approximately represented by an MPS.
  However, the recurrence relations of Chebyshev polynomials are broken by the approximation, leading to an error which is accumulated with the increase of the order of expansion.
  Here we propose a reorthonormalization approach to remove this error introduced in the loss of orthogonality of the Chebyshev polynomials.
  Our approach, as illustrated by comparison with the exact results for the one-dimensional XY and Heisenberg models, improves significantly the accuracy in the calculation of dynamical correlation functions.
\end{abstract}

\maketitle

\section{INTRODUCTION} \label{sec:intro}

Dynamical correlation functions, such as optical conductivity or
single-particle spectral function, are measurement quantities and are of central importance in condensed matter physics.
However, calculation of these quantities is a challenging theoretical problem.
During the last two decades, several approaches have been proposed to calculate spectral functions using the density matrix renormalization group (DMRG)~\cite{1992White} or other methods related to tensor network states~\cite{PhysRevLett.75.3537,PEPS,PhysRevX.4.011025}.
In 1987, Gagliano and Baliseiro~\cite{PhysRevLett.59.2999} (1987) proposed a continued-fraction method (also called the Lanczos vector method) to evaluate dynamic correlation functions based on the Lanczos diagonalization.
Their method was first adopted in the DMRG calculation by Hallberg~\cite{1995Hallberg} in 1995.
While this method requires only modest numerical resources, reliable results can be obtained only limited to low frequencies.
In 1989, a correction vector method was proposed by Soos and Ramasesha~\cite{ISI:A1989R859900047} to improve the continued-fraction method.
This correction vector method was adopted in the DMRG calculation first by Ramasesha \emph{et al.}~\cite{PhysRevB.55.8894} (1997) and later by K\"uhner and White~\cite{1999WhiteLDMRG} (1999).
The key idea is to take the Green's function at a particular frequency as a target state and calculate it by solving a set of large but sparse linear equations.
This method can generate very accurate result at any given frequency, but its computational cost is very high.
In 2002, Jeckelmann~\cite{PhysRevB.66.045114} improved the correction vector method and showed that it is much more efficient and accurate to determine the correction vector by minimizing a cost function.
Recently, Dargel~\emph{et al.} proposed an adaptive Lanczos vector method~\cite{2011Dargel} (2011) and an MPS-based Lanczos method~\cite{2012Dargel} (2012) that invokes the idea of orthogonalization to improve the accuracy of dynamic correlation functions.
More recently, Nocera and Alvarez~\cite{2016Nocera} (2016) proposed a Krylov-space approach to replace the conjugate gradient method in the calculation of correction vectors.

Dynamic quantities can be also calculated from the Fourier transformation of the time-dependent Green's function.
However, to obtain good frequency resolution, one has to calculate the correlation functions over a long time interval, which is limited by either a loss of accuracy due to the approximation used or by finite-size effects such as reflections from open ends.
In 2002, Cazalilla and Marston evaluated the time-dependent correlation functions for a one-dimensional (1D) system under an applied bias using the DMRG~\cite{2002Cazalilla}.
In 2003, Luo~\emph{et al.}~\cite{PhysRevLett.91.049701} proposed a pack-keeping approach  to optimize the basis states that are retained with the time evolution.
In 2004, Vidal introduced the so-call time-evolving block-decimation (TEBD) method to efficiently simulate time-evolution of 1D systems with short-range interactions using MPS~\cite{2004Vidal}.
Based on this approach, an adaptive time-dependent DMRG method was explored by Daley \emph{et al.}~\cite{2004Daley} and by White and Feiguin~\cite{PhysRevLett.93.076401}.

Long-time dynamics, however, remains challenging due to the linear growth of the entanglement entropy as the state evolves in time~\cite{2005Calabrese},
which implies that an exponentially growing bond dimension is required in
order to keep the error fixed.
In 2009, Ba\~{n}uls~\emph{et al.}~\cite{PhysRevLett.102.240603} proposed a folding scheme to present more efficiently the entanglement structure of the tensor-network state in the time direction, providing an accurate method for evaluating long-time dynamics of the ground state in the thermodynamic limit.
A quantum transfer-matrix DMRG extension of this method to finite temperatures was proposed by Huang~\emph{et al.}~\cite{PhysRevB.89.201102}.
Generalized DMRG approaches have also been extended to simulate time evolution at finite temperatures by introducing auxiliary degrees of freedom to
purify the thermal statistical operators~\cite{2009Barthel,PhysRevLett.108.227206}.

Recently, Holzner \emph{et al.}~\cite{2011Holzner} (2011) proposed a Chebyshev matrix-product-states (CheMPS) method to calculate dynamical spectral functions for operators $A$ and $B$ at zero temperature
\begin{equation} \label{eq:spec}
  G_{AB}(\omega )=\langle 0|A\delta (\omega -H+E_{0})B|0\rangle ,
\end{equation}
where $H$ is the Hamiltonian, $E_0$ is the ground state energy and $|0\rangle$ is the corresponding eigenstate.
The main idea is to use the Chebyshev polynomials to expand the delta function and to use a set of MPS to represent the Chebyshev vectors generated in the Chebyshev expansion.
This method offers precise and convenient control of the accuracy and resolution.
It yields results with accuracies comparable to those of the correction-vector DMRG, but at dramatically reduced numerical cost.

The CheMPS provides a balanced scheme between cost and accuracy. It resolves spectral functions well if the spectral weight is concentrated or limited to low energy part, but cannot resolve the high-energy spectrum accurately, similar as in the continued-fraction DMRG.
Since each Chebyshev vector is approximated by an MPS at each step of the Chebyshev expansion, the recurrence relation between different Chebyshev vectors is satisfied only approximately.
Thus the resolution may not be truly improved with the increase of the series number of Chebyshev expansion.

In this paper, we introduce a reorthonormalization scheme to improve the accuracy in the calculation of spectral function based on the CheMPS. (Similar approach was used to improve the Lanczos-MPS calculation~\cite{2012Dargel}).
In the discussion below, for simplicity, we refer our method as ReCheMPS.
In this method, the Chebyshev expansion is applied twice.
We first carry out a CheMPS calculation to obtain a set of Chebyshev vectors represented using MPS, and then reorthonormalize these MPS to obtain a set of many-body basis states~\cite{2016Huang}.
Within the truncated Hilbert space spanned by this set of basis states, we re-diagonalize the Hamiltonian and evaluate the spectral function, using the Chebyshev expansion again.
This method, as shown below, improves the accuracy of CheMPS significantly, with very limited increase of computational cost in comparison with CheMPS.
Furthermore, we propose a simple approach to extrapolate the finite-size results to the thermodynamic limit using the eigen-energies of the truncated Hamiltonian and the corresponding spectral weights.
By comparison with the exact results in the thermodynamic limit, we find that this approach offers an efficient way to perform the finite-size scaling in the entire energy range.

This paper is organized as follows.
In Sec.~\ref{sec:method}, after a brief review of the CheMPS in Sec.~\ref{Sec:CheMPS}, the ReCheMPS is introduced in Sec.~\ref{Sec:ReCheMPS}. We propose a finite size scaling approach in Sec~\ref{sec:method-scale}, based on the eigen-spectra and the spectral weights obtained from the effective Hamiltonian in the orgonalized but truncated basis space.
Then we calculate the magnetic structure factor using the ReCheMPS for the 1D XY and Heisenberg models.
The results are presented in Sec.~\ref{sec:xy} for the XY model and in Sec.~\ref{sec:hsg} for the Heisenberg model, and compared with those obtained by the CheMPS.
The finite size scaling result is showen in Sec.~\ref{sec:scale} and compared with the exact results obtained in the thermodynamic limit.
Finally, Sec.~\ref{sec:conclusion} gives a summary.

\section{Method} \label{sec:method}
\subsection{The CheMPS method} \label{Sec:CheMPS}

To carry out a CheMPS calculation~\cite{2011Holzner}, one needs to first calculate the wave function of the ground state $|0\rangle$ and the corresponding energy $E_0$ using the DMRG or variational MPS methods.
In order to determine the spectral function $G_{AB}(\omega )$ defined in Eq.~(\ref{eq:spec}) in a frequency interval $[0,W]$, we rescale and shift the frequency and the Hamiltonian such that the nonzero parts of spectral function are completely concentrated on the frequency interval $[-W^{\prime },W^{\prime }]$, where $W^{\prime }$ is usually taken to be slightly smaller than 1 for numerical stability~\cite%
{2011Holzner}. Then the spectral function becomes
\begin{equation}
G_{AB}(\omega ^{\prime })=\frac{1}{a}\langle 0|A\delta (\omega ^{\prime
}-H^{\prime })B|0\rangle  \label{eq:spec2}
\end{equation}%
where $a=W/(2W^{\prime })$, and
\begin{equation} \label{eq:rescaled_Hw}
\omega ^{\prime } = \frac{\omega }{a}-W^{\prime },
\qquad H^{\prime } = \frac{H-E_{0}}{a}-W^{\prime }.
\end{equation}

The $\delta $-function in Eq.~(\ref{eq:spec2}) can be expanded by an
infinite Chebyshev polynomial series~\cite{2006Weisse}. In practical
calculation, this expansion is approximated by the first $N$ terms of the
Chebyshev polynomials~\cite{2011Holzner}, i.e.
\begin{equation}
\delta (\omega ^{\prime }-H^{\prime })= \frac{1}{b} \left[ g_{0}+2\sum_{n=1}^{N-1}g_{n}T_{n}(H^{\prime })T_{n}(\omega ^{\prime })\right] ,
\label{Eq:delta}
\end{equation}
where $b=\pi \sqrt{1-\omega ^{\prime 2}}$ and $g_{n}$ are the damping factors to suppress the Gibbs oscillations introduced by the truncation to the infinite Chebyshev
series. There are several different choices for these damping factors~\cite{2006Weisse}, here we use the Jackson damping defined by
\begin{eqnarray}
g_{n} &=&\frac{1}{N+1}(N-n+1)\cos {\frac{\pi n}{N+1}}  \nonumber \\
&&+\frac{1}{N+1}\sin {\frac{\pi n}{N+1}}\cot {\frac{\pi }{N+1}.}
\end{eqnarray}

In Eq.~(\ref{Eq:delta}), $T_{n}(x)$ are the Chebyshev polynomials defined by the recurrence relation
\begin{equation}
T_{n+1}(x)=2xT_{n}(x)-T_{n-1}(x)  \label{eq:Tn_relations}
\end{equation}
with $T_{0}(x)=1$ and $T_{1}(x)=x$. $T_{n}(x)$ can be also represented as
\begin{equation}
T_{n}(x)=\cos \left( n\,\cos ^{-1}x\right) =\cosh \left( n\cosh
^{-1}x\right) .
\end{equation}

Substituting Eq.~(\ref{Eq:delta}) into Eq.~(\ref{eq:spec2}), we obtain
\begin{equation}
G_{AB}(\omega ^{\prime })=\frac{1}{ab}\left[
g_{0}\mu _{0}+2\sum_{n=1}^{N-1}g_{n}\mu _{n}T_{n}\left( \omega ^{\prime
}\right) \right] ,  \label{eq:Gw_expansion}
\end{equation}%
where $\mu _{n}$ are the Chebyshev moments defined by
\begin{equation}
\mu _{n}=\langle 0|AT_{n}(H^{\prime })B|0\rangle.
\end{equation}
To obtain the Chebyshev moments $\mu _{n}$, we first calculate the Chebyshev
vectors
\begin{equation}
|\psi _{n}\rangle =T_{n}(H^{\prime })B|0\rangle .
\end{equation}
From the recurrence relation Eq.~(\ref{eq:Tn_relations}), we find that
$ |\psi _{0}\rangle = B|0\rangle $, $|\psi _{1}\rangle =H^{\prime }|\psi _{0}\rangle $, and
\begin{equation}
   |\psi _{n}\rangle = 2H^{\prime }|\psi _{n-1}\rangle -|\psi _{n-2}\rangle ,
\end{equation}
These Chebyshev vectors $|\psi _{n}\rangle $ span a \textit{Krylov
basis space. }

In the CheMPS calculation, the ground state $|0\rangle $ and all the Chebyshev
vectors $|\psi _{n}\rangle $ are approximately represented by MPS. In this
case, the above equations are valid just approximately.
The MPS representation of the first two Chebyshev vectors are determined by variationally minimizing the functions
\begin{eqnarray}
f_{0}\left( \psi _{0}\right) &=&\left\Vert |\psi _{0}\rangle -B|0\rangle
\right\Vert ^{2},  \label{Eq:CostFunc1} \\
f_{1}\left( \psi _{1}\right) &=&\left\Vert |\psi _{1}\rangle -H^{\prime
}|\psi _{0}\rangle \right\Vert ^{2}.
\end{eqnarray}
Similarly, other Chebyshev vectors are determined by minimizing the cost
function%
\begin{equation}  \label{Eq:CostFunc3}
f_{n}\left( \psi _{n}\right) =\left\Vert |\psi _{n}\rangle -2H^{\prime
}|\psi _{n-1}\rangle +|\psi _{n-2}\rangle \right\Vert ^{2}.
\end{equation}
From these Chebyshev vectors, the Chebyshev moments are determined by the formula
\begin{equation}  \label{Eq:Moments}
\mu _{n}=\langle 0|A|\psi _{n}\rangle .
\end{equation}

Given a Chebyshev expansion number $N$, it is known that the CheMPS has a spectral resolution of order $O(W/N)$ in the interval $\omega \in \lbrack 0,W]$~\cite{2011Holzner}. When the bandwidth is broadened or the spectrum changes drastically, one has to increase $N$ to increase the resolution.

\subsection{The ReCheMPS Method}
\label{Sec:ReCheMPS}

Two approximations are used in the CheMPS.
The first is to truncate the infinite Chebyshev series and approximate it by a finite polynomial. The error such introduced can be reduced by increasing the number of Chebyshev polynominals $N$.
The second is to approximate each Chebyshev vector by an MPS. This is a more severe approximation because it breaks the orthonormal condition of Chebyshev series, and the error is accumulated in the calculation of Chebyshew vectors with Eq.~(\ref{Eq:CostFunc3}).

Here we propose a reorthonormalization scheme to solve the second problem.
We start from the Chebyshev vectors obtained using the CheMPS, and then carry out the following reorthonormalization calculation:
\begin{enumerate}

\item Reorthonormalize the Chebyshev vectors $|\psi_\alpha\rangle$ using the Gram-Schmidt orthogonalization process
\begin{equation}
\left\vert \Psi _{\alpha }\right\rangle =c_{\alpha }\left( 1-\sum_{\beta
<\alpha }\left\vert \Psi _{\beta }\right\rangle \left\langle \Psi _{\beta
}\right\vert \right) \left\vert \psi _{\alpha }\right\rangle,
\label{Eq:NewBasis}
\end{equation}%
so that
\begin{equation}
\left\langle \Psi _{\alpha }|\Psi _{\beta }\right\rangle =\delta _{\alpha
\beta }.
\end{equation}
Here $\left\vert \Psi _{\alpha }\right\rangle $ with $\left\vert \Psi
_{0}\right\rangle = B\left\vert0\right\rangle $ are the new orthonormalized
basis states. $c_{\alpha }$ are the normalization constants.

\item Represent the Hamiltonian and other operators in the basis space
spanned by $\left\{ \left\vert \Psi _{\alpha }\right\rangle ,\alpha
=0,...,N-1\right\} $,
\begin{equation}
H_{\alpha \beta }^{\prime \prime  } = \langle \Psi _{\alpha }|H^{\prime
}|\Psi _{\beta }\rangle .   \label{Eq:H-matrix-elements}
\end{equation}

\item Redo the Chebyshev expansion and generate $N_r$ new Chebyshev vectors $|\psi_\alpha{^{\prime \prime}}\rangle$ using $H^{\prime \prime  }$ in this new basis space.

\item Find the new Chebyshev moments $\mu _{n}$ using Eq.~(\ref{Eq:Moments}) and
the spectral function using Eq.~(\ref{eq:Gw_expansion}).
\end{enumerate}

In this new scheme, the Chebyshev expansions are applied twice.
In the first expansion, $N$ approximate Chebyshev vectors represented by MPS
are evaluated by variationally minimizing the cost functions defined in
Eqs.~(\ref{Eq:CostFunc1} -~\ref{Eq:CostFunc3}). These MPS are not
orthonormalized due to the truncation in the bond dimension.

In the second expansion, the Hamiltonian is approximately replaced by $H^{\prime\prime}$, which is an $N\times N$ matrix defined in the space spanned by $\{ |\Psi_\alpha \rangle , \, (\alpha = 0 \cdots N-1)\}$, and $N_r$ Chebyshev vectors that satisfy the standard recurrence relations of Chebyshev polynominals are generated using $H^{\prime\prime}$.
Here $N_r$ is the number of Chebyshev vectors used for approximating the $\delta$-function using Eq.~(\ref{Eq:delta}) for the Hamiltonian $H^{\prime\prime}$.
It can be set either equal to $N$ or as an independent control parameter.
The Chebyshev expansion Eq.~(\ref{Eq:delta}) becomes exact in the limit $N_r\rightarrow \infty$.

In the calculation of ReCheMPS, only the inner products between any two MPS already generated in the CheMPS need to be calculated.
Its cost in both computational time and memory is quite small in comparison with the calculation of CheMPS. Thus, as shown below, our reorthogonalization scheme can improve significantly the accuracy of the CheMPS, but with very limited increase in the computational time.

The Hamiltonian $H^{\prime\prime}$ has a finite dimension. We can diagonalize it to obtain its spectrum
\begin{equation} \label{Eq:Hpp}
  H^{\prime\prime} | n \rangle = \lambda_n |n\rangle \qquad (n = 1 \cdots N),
\end{equation}
where $\lambda_n$ and $|n\rangle$ are the eigen-pairs of $H^{\prime\prime}$, which are arranged in an ascending order, \emph{i.e.} $\lambda_n <\lambda_m$ if $n< m$.
In this truncated Hilbert space, the spectral function is given by
\begin{equation} \label{Eq:delta2}
   G_{AB} (\omega ) = \sum_{n=1}^N \langle 0 | A | n\rangle \langle n | B |0 \rangle \delta ( \omega - \lambda_n + E_0 ) .
\end{equation}

In the second Chebyshev expansion, the above delta-functions are blurred by $N_r$ Chebyshev polynomials.
In this round of expansion, there is no technical difficulty to take arbitrary many Chebyshev vectors in the expansion.
Therefore, we can set $N_r$ equal or not equal to $N$, depending on the resolution required. Eq.~(\ref{Eq:delta2}) is recovered in the limit $N_r \rightarrow \infty$.

Furthermore, as the spectrum of $H^{\prime\prime}$ is known, we can always set the bandwidth of $H^{\prime\prime}$, represented by $W_*$, exactly equal to the difference between the largest and smallest eigenvalues of $H^{\prime\prime}$, {\it i.e.} $W_* =\lambda_N - \lambda_1$, and find all the Chebyshev vectors without taking any approximation in the second Chebyshev expansion.
Clearly, $W_* < W$, thus if the same number of Chebyshev vectors is used as in CheMPS, {\it i.e.} setting $N_r = N$, the energy resolution, which is qualitatively determined by the ratio $W_*/N_r$, should be higher in ReCheMPS.

\subsection{\label{sec:method-scale} Finite-size scaling}

The CheMPS offers a convenient way to control the accuracy and resolution of the spectral function by simply adjusting the expansion order $N$.
This is an advantage that can be used to analyze the finite-size effects~\cite{2011Holzner}.
For example, one can mimic the infinite lattice limit ($L\rightarrow \infty$) by
choosing a small enough $N$ or more precisely a large $W/N$ to smear out the finite-size subpeaks. However, in practical calculation, it is difficult to adjust $N$ to reduce the finite size effect for the systems with different lattice lengths.

An alternative approach is to take a large $N$ limit and assume that all spectral peaks can be reliably resolved. By fitting the peaks by Gaussian functions, one can obtain the central position and the weight for each peak. The spectral function at each peak position is then defined by the ratio between the peak weight and the half distance between the two neighboring peaks.
This approach does yield quite impressive results for the Heisenberg model~\cite{2011Holzner}.
However, it relies on the accurate calculation of spectral peaks, which is in fact difficult due to the MPS approximation of Chebyshev vectors.
Technically, it is also difficult to use a large $N$ to do the Chebyshev expansion when the lattice size $L$ or the bond dimension $D$ becomes large.

Here we propose a novel approach to perform the finite-size scaling.
In Sec.~\ref{Sec:ReCheMPS}, we show that by diagonalizing the Hamiltonian $H^{\prime \prime}$ in the truncated basis space, we can approximately represent the spectral function as
\begin{equation}
  S(k,\omega ) = \sum_{n} \delta (\omega - \lambda_{n} +E_{0}) W_n ,
\end{equation}
where $\lambda_n$ and $|n\rangle$ are the eigen-pairs of $H^{\prime\prime}$, defined in Eq.~(\ref{Eq:Hpp}), and
\begin{equation}
  W_n = \langle 0|S_{-k}^z |n\rangle \langle n| S_k^Z|0\rangle
\end{equation}
is the spectral weight corresponding to $\lambda_n$.

Let us denote $a_n$ as the midpoint between $\lambda_{n-1}$ and $\lambda_n$, i.e.
\begin{equation}
  a_n = \frac{\lambda_{n-1}+\lambda_{n}}{2} , \qquad (1<n\le N) .
\end{equation}
The average spectral weight within the interval between $a_n$ and $a_{n+l}$ ($l>0$) is given by
\begin{equation} \label{Eq:SpecFunc}
  S(k, \bar{\omega}_{n,l}) = \frac{1}{\Delta_{n,l}} \sum_{0\le m < l} W_{n+m},
\end{equation}
where
\begin{equation}
  \Delta_{n,l} = a_{n+l} - a_n
\end{equation}
is the energy difference between $a_{n+l}$ and $a_n$, and
\begin{equation}
  \bar{\omega}_{n,l} = \frac{a_n + a_{n+l}}{2}
\end{equation}
is the average energy of these two points.
Here $l$ should be taken such that the interval between $a_n$ and $a_{n+l}$ contains just one main peak. 

\section{Results}
\label{Sec:Result}

\begin{figure}[t]
\centering
\includegraphics[width=8.0cm]{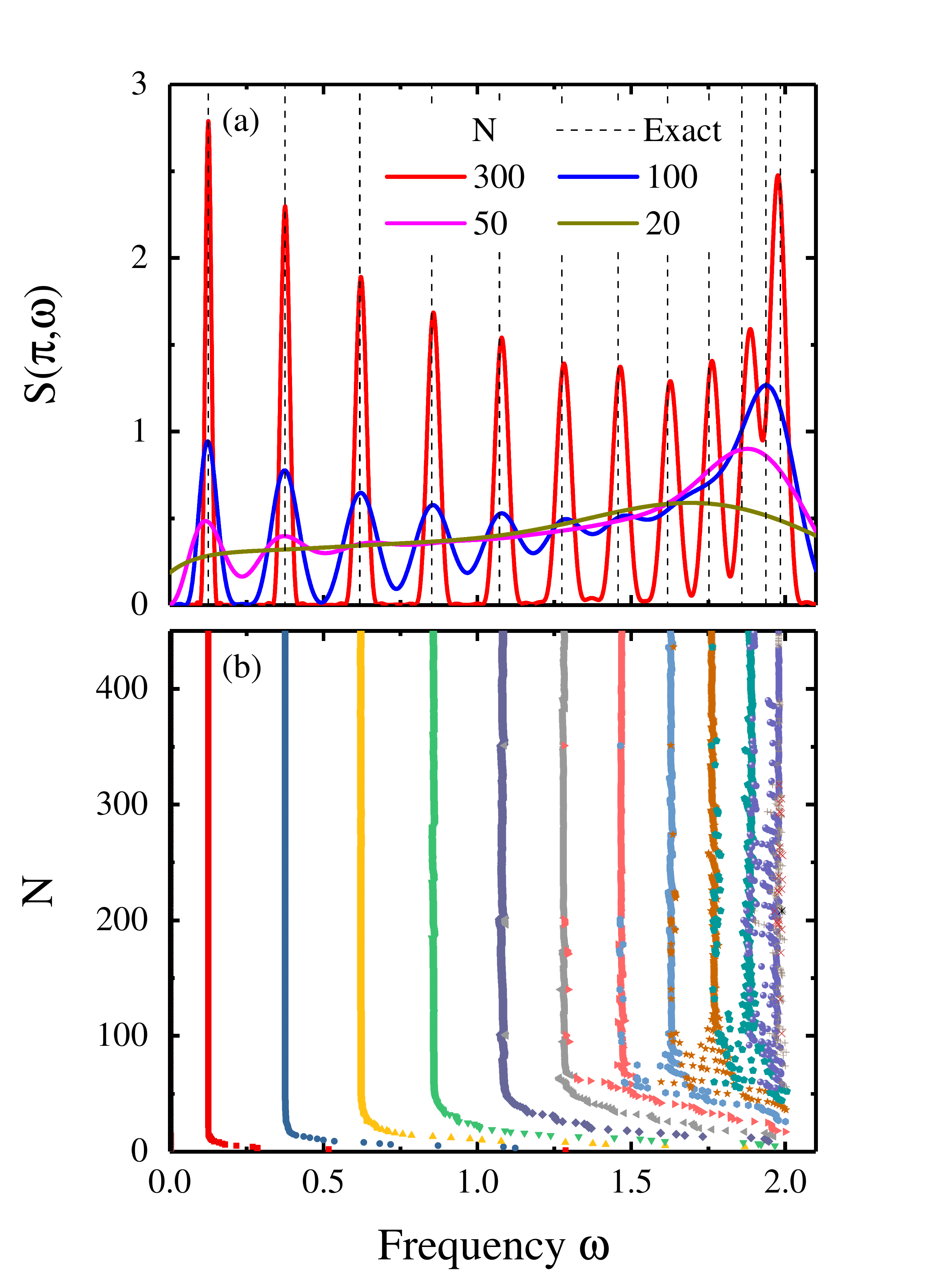}
 \caption{
 (a) Spin structure factor $S(k=\pi,\omega)$ and (b) the peak frequencies of $S(\pi, \omega)$ for the 1D XY model obtained using the ReCheMPS with $L=24$, $D=32$ and varying parameter $N$.
 The exact excitation energies of the model are represented by the dashed lines.
 }
\label{fig:XY_L24_comb}
\end{figure}

\begin{figure}[t]
\centering
\includegraphics[width=8.0cm]{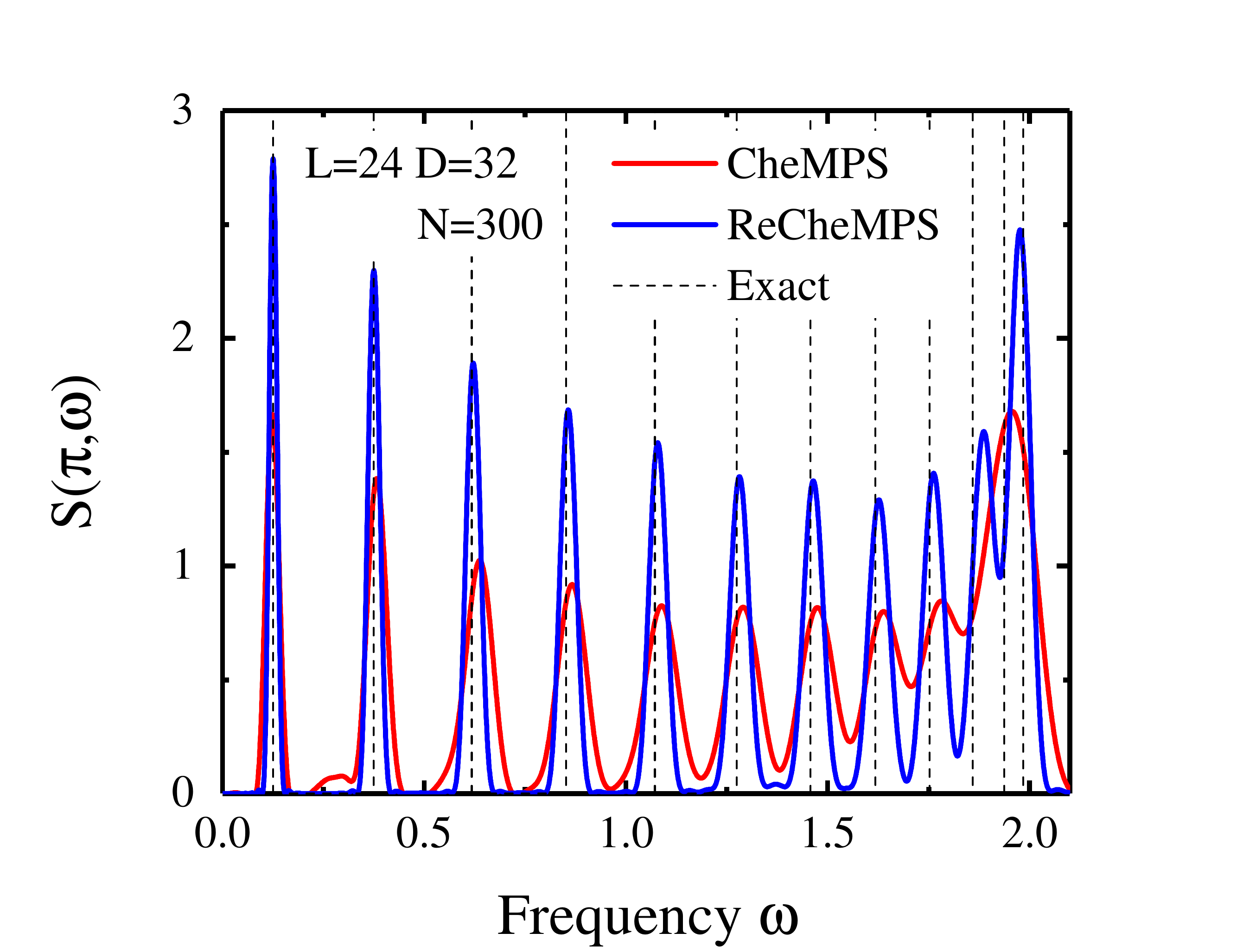}
 \caption{Comparison between the results of $S(\pi,\omega)$ obtained using CheMPS and ReCheMPS for the 1D XY model with $L = 24$, $D =32$, and $N =N_r= 300$. The exact excitation energies are represented by the dashed lines.
}
\label{fig:L24xy}
\end{figure}

In this section, we present the results obtained by the ReCheMPS and compare them with those obtained by the CheMPS for both the $S=1/2$ XY and the $S=1/2$ antiferromagnetic Heisenberg models in one dimension.
These models are respectively defined by the Hamiltonians
\begin{equation}
H_{XY}=-\sum_{r=1}^{L-1}\left( S_{r}^{x}S_{r+1}^{x}+S_{r}^{y}S_{r+1}^{y}\right) ,
\label{f:Hxy}
\end{equation}
and
\begin{equation}
H_{Heis}= \sum_{r=1}^{L-1}\mathbf{S}_{r}\cdot \mathbf{S}_{r+1},
\end{equation}
where $\mathbf{S}_{r}=(S_r^x, S_r^y, S_r^z)$ is the spin operator and $L$ is the lattice length.
For both cases, open boundary conditions are assumed.

We have calculated the spin-spin correlation function defined by
\begin{equation}
  S (k, \omega ) = \langle 0 | S_{-k}^z \delta (\omega - H + E_0) S_k^z | 0 \rangle ,
\end{equation}
where $S_k^z$ is the spin operator defined in momentum space
\begin{equation}
  S^z_{k}=\sqrt{\frac{2}{L+1}}\sum_{r=1}^{L}\mathrm{sin}(kr)S^z_{r},  \label{f:Sq}
\end{equation}
and $k=n\pi /(L+1)$ $(n=1, ... ,L)$ is the quasi-momentum.

In the calculation of this correlation function, the energy spectrum is bounded by the band width $W$, which equals $2$ and $\pi$ for the 1D XY and Heisenberg models,  respectively.
However, in the CheMPS calculation, this energy bound is broken by the approximation used in the MPS representation of Chebyshev vectors, and the effective band width can be larger than the exact one.
To avoid this overflow of energy, we set $W$ equal to a value which is greater than the exact one so that the energy spectrum can be fully covered in the first Chebyshev expansion.
In this work, we take $W=10$ for both models.
By doing this, we do not need to employ the energy truncation method used in Ref.~[\onlinecite{2011Holzner}].

During the second Chebyshev expansion, as already mentioned in Sec.~\ref{Sec:ReCheMPS}, we take the bandwidth exactly equal to the difference between the largest and the smallest eigen-energies, {\it i.e.} $W_*$, of the approximate Hamiltonian $H^{\prime\prime}$.
For all the calculation we have done, we find that $W_*<10$ and varies slightly with $N$.
The results presented below are obtained by taking $N_r = N$ if not particularly specified.

\subsection{\label{sec:xy} The XY-model}

\begin{figure}[t]
\centering
\includegraphics[width=8.0cm]{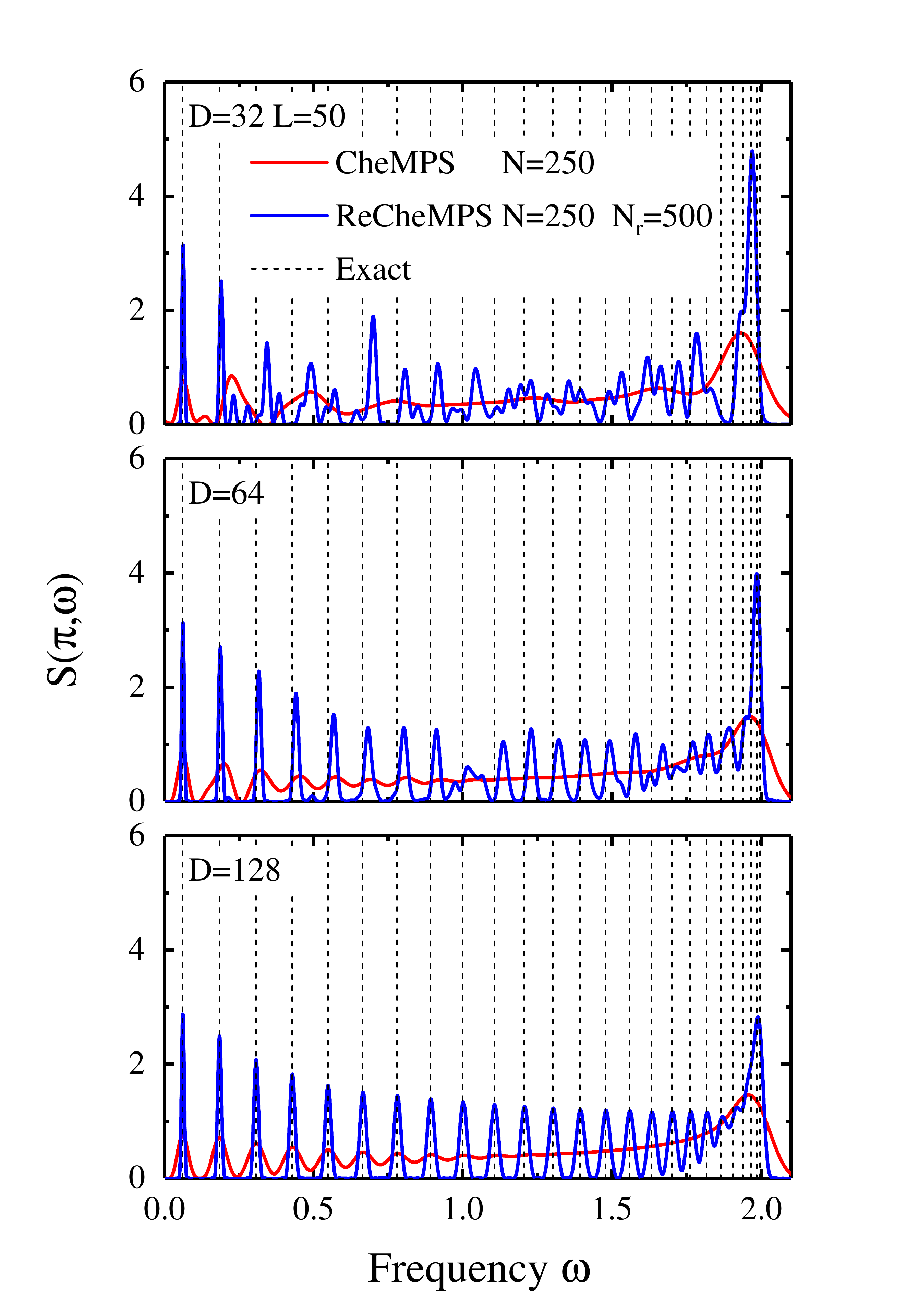}
  \caption{
  Comparison of the frequency dependence of $S(\pi, \omega)$ obtained by the ReCheMPS with that by the CheMPS at three different $D$ for the 1D XY model with $L=50$, $N=250$ and $N_r=500$.
  The exact excitation energies of the model are represented by the dashed lines.
  }
\label{fig5:L50xydisp}
\end{figure}

The 1D XY model can be mapped onto a free spinless fermion model through the Jordan-Wigner transformation~\cite{JWtrans}.
The spin-spin correlation function $S (k,\omega )$ contributes mainly by the
particle-hole excitations of spinless fermions, whose energy is bounded respectively from below and above by
\begin{equation}
\varepsilon_{XY}^{\min }(k) =|\mathrm{sin}k|, \qquad \varepsilon_{XY}^{\max }(k) =2|\mathrm{sin}\frac{k}{2}|.
\end{equation}
As $k\rightarrow \pi $, the spin-spin correlation function can be exactly
represented as
\begin{equation}
S(\pi ,\omega )=\sum_{k}\delta (\omega -\Omega _{k}),
\label{eq:Szz}
\end{equation}
where $\Omega _{k}=2 |\mathrm{cos}k|$. In the limit $L\rightarrow \infty $, it becomes
\begin{equation}
S(\pi ,\omega )=\frac{2}{\pi \sqrt{4-\omega ^{2}}}.  \label{eq:inf}
\end{equation}

Figure~\ref{fig:XY_L24_comb}(a) shows $S(k=\pi ,\omega )$ as a function of frequency $\omega$ for the 1D XY model obtained with ReCheMPS on the $L=24$ lattice.
As expected, the spectral resolution is improved with the increase of $N$.
The peak positions of $S(\pi ,\omega )$, as illustrated in Fig.~\ref{fig:XY_L24_comb}(b), converge rapidly to the
exact energies, $\Omega_k = 2 |\cos k|$, again with the increase of $N$.

Figure~\ref{fig:L24xy} compares the result obtained by ReCheMPS with that by CheMPS.
The improvement of ReCheMPS over CheMPS is quite significant. By keeping the same number of Chebyshev vectors, the spectral resolution of ReCheMPS is clearly better than that of CheMPS. Besides, the ReCheMPS can also improve the accuracy of the peak positions, in comparison with the exact ones, especially for the intermediate and high energy peaks.

Moreover, this improvement becomes more pronounced when the system size becomes larger.
Fig.~\ref{fig5:L50xydisp} shows how the spectral function $S(\pi , \omega)$ changes with $D$ for the system of $L=50$.
By comparison with the result shown in Fig.~\ref{fig:L24xy}, we find that a larger $D$ is needed in order to obtain the same accuracy in the spectral function when the lattice size is increased, while the result obtained by the CheMPS does not show significant improvement with the increase of D because of the violation of the orthonormal condition of Chebyshev series.
The peak positions we obtain with the ReCheMPS agree well with the exact ones, especially for the case $D=128$.

\begin{figure}[t]
\centering
\includegraphics[width=8.0cm]{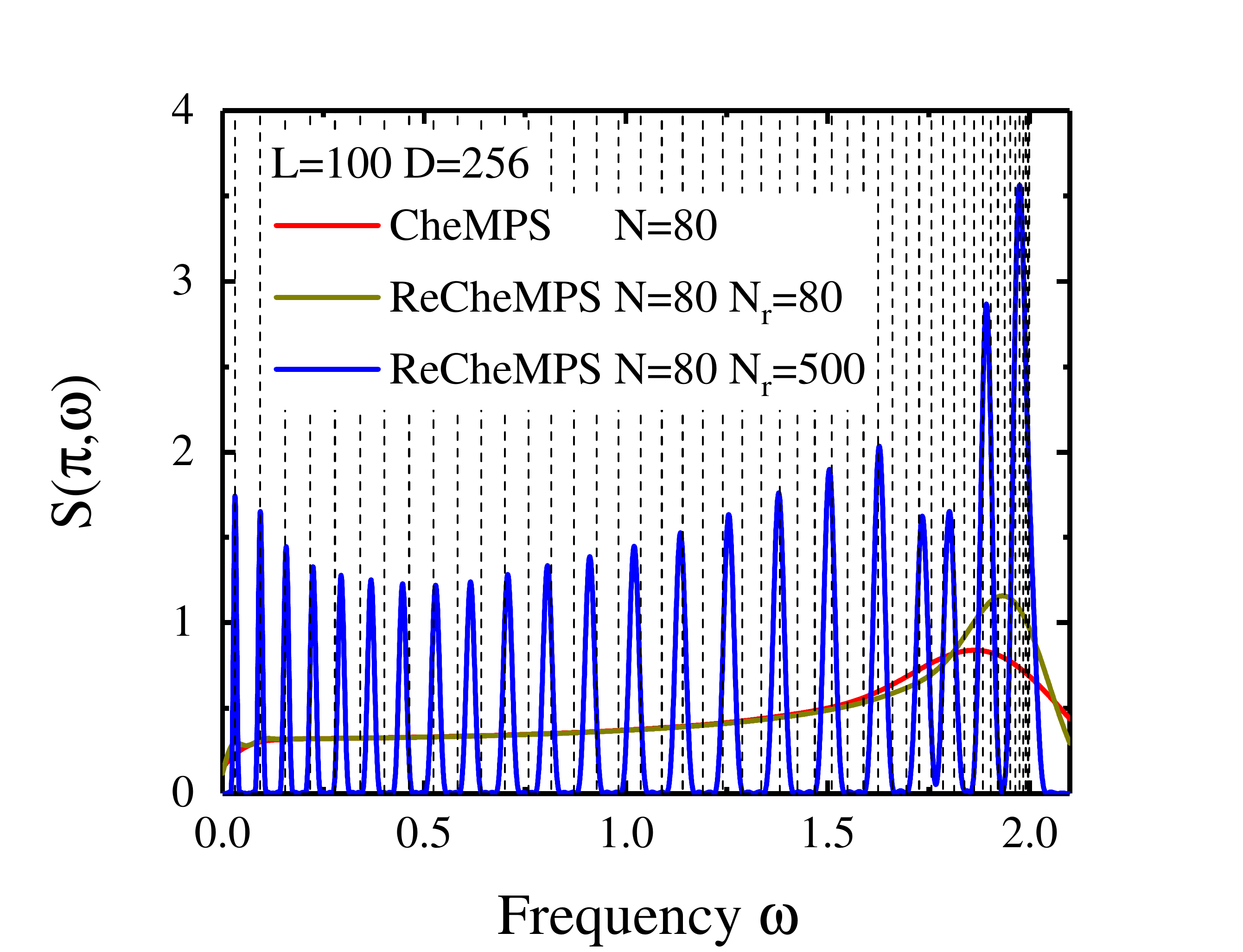}
  \caption{Frequency dependence of the spectral function $S(\pi, \omega)$  obtained
using CheMPS and ReCheMPS of the 1D XY model with $L=100$ and $D=256$, while $N$ and $N_r$ listed in figure.
  The exact excitation energies of the model are represented by the dashed lines.
}
\label{fig:L100xybig}
\end{figure}

With the further increase of the lattice size, the computational cost (including both computation time and memory) increases rapidly with increasing $N$, and it is difficult to use a large $N$ to do the Chebyshev expansion.
This limits the resolution of the CheMPS calculation.
However, in the ReCheMPS calculation, we can use a large $N_r$ to increase the resolution.
Fig.~\ref{fig:L100xybig} shows, as an example, how the resolution changes with the increase of $N_r$ in ReCheMPS for the XY model with $L=100$ and $N=80$.
When $N_r=N=80$, the resolutions of both ReCheMPS and CheMPS are very poor. However, when we increase $N_r$ to 500, the resolution is greatly improved and the ReCheMPS produces all the low energy peaks accurately.

\begin{figure}[t]
\begin{center}
\includegraphics[width=8.0cm]{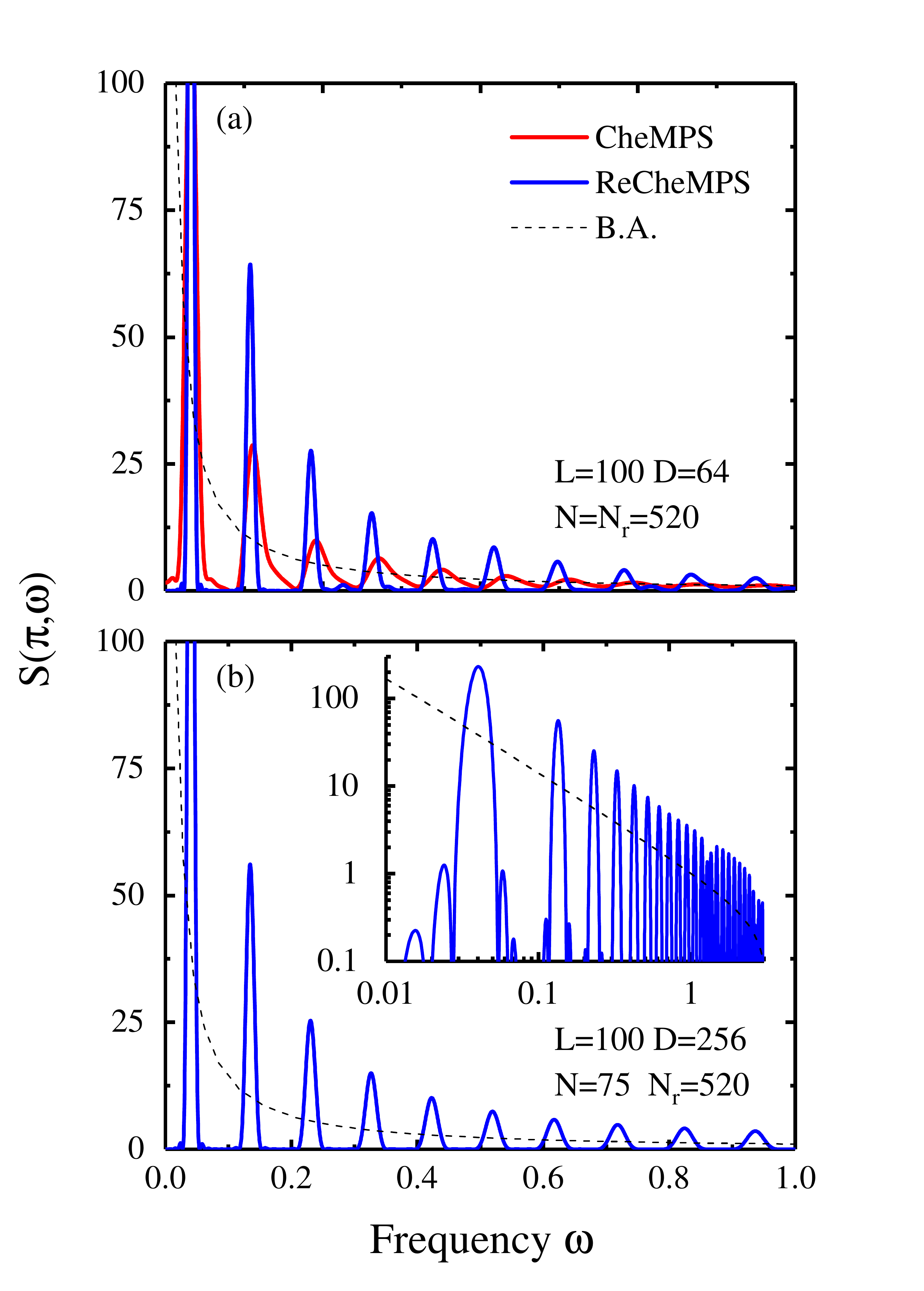}
  \caption{$S(\pi, \omega)$ of the 1D Heisenberg model of $L=100$. (a) Comparison between the results obtained by the ReCheMPS and the CheMPS with $D=64$ and $N=N_r=520$. 
  (b) The ReCheMPS result of $S(\pi , \omega)$ obtained with $D=256$ and $N=75$. The inset is in the logarithmic scales for both horizontal and vertical axes.
  The dash line denotes the result of Bethe Ansatz (B.A.) obtained in the thermodynamic limit $L\rightarrow \infty$.
} \label{fig:HeisenbergL100}
\end{center}
\end{figure}

\subsection{\label{sec:hsg} The Heisenberg model}

\begin{figure}[t]
\centering
\includegraphics[width=8.0cm]{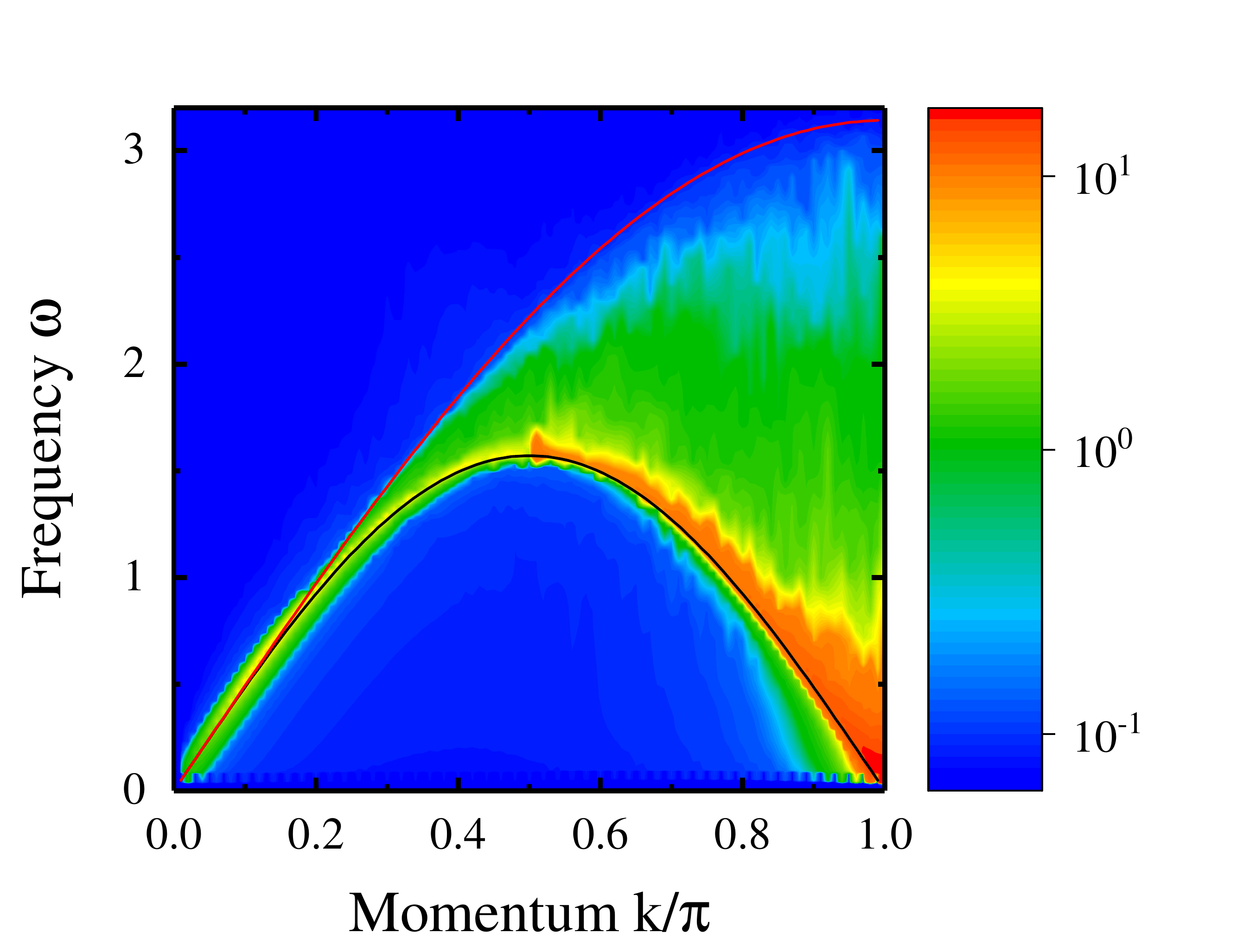}
  \caption{Intensity plot of $S(k, \omega)$ obtained by the ReCheMPS for the 1D Heisenberg model of $L=100$. $D=64$ and $N=300, N_r=1000$ are used. The two solid lines represent the lower and upper bounds stemmed from the two-spinon excitations~\cite{2011Holzner}, i.e. $\omega^{\min}_{\mathrm{Heis}}$ and $\omega^{\max}_{\mathrm{Heis}}$, defined in Eq.~(\ref{Eq:HeisMin}).}
\label{fig:peakHsgDiag}
\end{figure}

For the spin-1/2 Heisenberg model, the asymptotic behavior of the
dynamical spin structure factor is known from the Bethe ansatz solution~\cite
{2006Bethe}. It diverges at $k=\pi$ in the limit $\omega\rightarrow 0$ and $L\rightarrow \infty$ 
\begin{equation} \label{Eq:Heisenber1}
S(\pi ,\omega )\propto \frac{1}{\omega }\sqrt{\ln \frac{1}{\omega }} .
\end{equation}
Since the spectra weight of the Heisenberg model diverges in the zero energy limit,
we find that the low energy spectrum can be accurately calculated with relatively small $D$ in a large lattice system.

Figure~\ref{fig:HeisenbergL100}(a) compares the results obtained using CheMPS and ReCheMPS with $D=64$ for the Heisenberg model of $L=100$.
The spectral peak obtained with the ReCheMPS is higher because the ReCheMPS has higher resolution than the CheMPS.
Moreover, the CheMPS peaks drop quickly with the increase of energy and become almost invisible at the high energy end.
But the high energy peaks obtained with the ReCheMPS can be seen in the whole energy range.

The peak energies obtained with these two methods agree well with each other in the low-energy limit.
But the difference becomes larger when the energy is increased.
Similar as for the XY model, we find that the low energy spectral peaks obtained with  ReCheMPS converge quickly with the increase of $D$.
The accuracy in the high-energy spectral function,  as shown in Fig.~\ref{fig:HeisenbergL100}(b),can be improved by increasing $D$ from 64 to 256.
As expected, the spectral weight drops quickly with increasing energy.
The energy dependence of the peak height, which can be more clearly seen from the log-log plot of $S(\pi , \omega )$ shown in the inset of Fig.~\ref{fig:HeisenbergL100}(b), follows the trend that is expected from the Bathe Ansatz solution Eq.~(\ref{Eq:Heisenber1}) obtained in the thermodynamic limit.

We also calculate the spin structure factor $S(k,\omega)$ for the 1D Heisenberg model in the whole momentum space.
An intensity plot of $S(k,\omega)$ is shown in Fig.~\ref{fig:peakHsgDiag}.
$S(k,\omega)$ can be measured by inelastic neutron scattering spectroscopy.
Our ReCheMPS calculation produces correctly the behavior of spinon continuum spectra.
The lower and upper bounds of the continuum agree well
the exact results, i.e.
\begin{equation}
  \omega^{\min}_{\mathrm{Heis}} = \frac{\pi}{2}|\sin k|, \qquad
  \omega^{\max}_{\mathrm{Heis}}  =  \pi|\sin \frac{k}{2}|. \label{Eq:HeisMin}
\end{equation}

\subsection{\label{sec:scale} Finite-size scaling}

Here we use the method introduced in Sec.~\ref{sec:method-scale} to perform a finite size scaling analysis for the spectra obtained using the effective Hamiltonian $H^{\prime \prime}$.
In practical calculation of the spectral function using Eq.~(\ref{Eq:SpecFunc}), the energy interval $\Delta_{n,l}$ should be taken such that it is roughly equal to the energy difference between two neighboring main peaks in the spectral function.
For example, for the spectral shown in Fig.~\ref{fig:HeisenbergL100}(b), we can clearly identify ten peaks in the energy interval from 0 to 1.
In some cases, it is sufficient just to take $l=1$.
But in the following cases, one should enlarge the value of $l$ so that $\Delta_{n,l}$ contains more than one energy level:
(1) two or more energy levels are nearly degenerate, such that their energy difference is significantly smaller than the main peak distance;
and (2) add one more energy level to $\Delta_{n,l}$ does not change significantly the total spectral weight within this enlarged energy interval.
Practically, one can start with $l=1$, and then add one or more energy level in.
If the change in the total spectral weight within the energy interval before and after adding the energy levels is smaller than a threshold, one can then augment $l$ to contain all those energy levels, provided that $\Delta_{n,l}$ is not significantly larger than the peak distance.
The threshold can be set, for example, from 0.1 to a few percent of the total spectral weight depending on the accuracy of calculated data.

\begin{figure}[t]
\centering
\includegraphics[width=8.0cm]{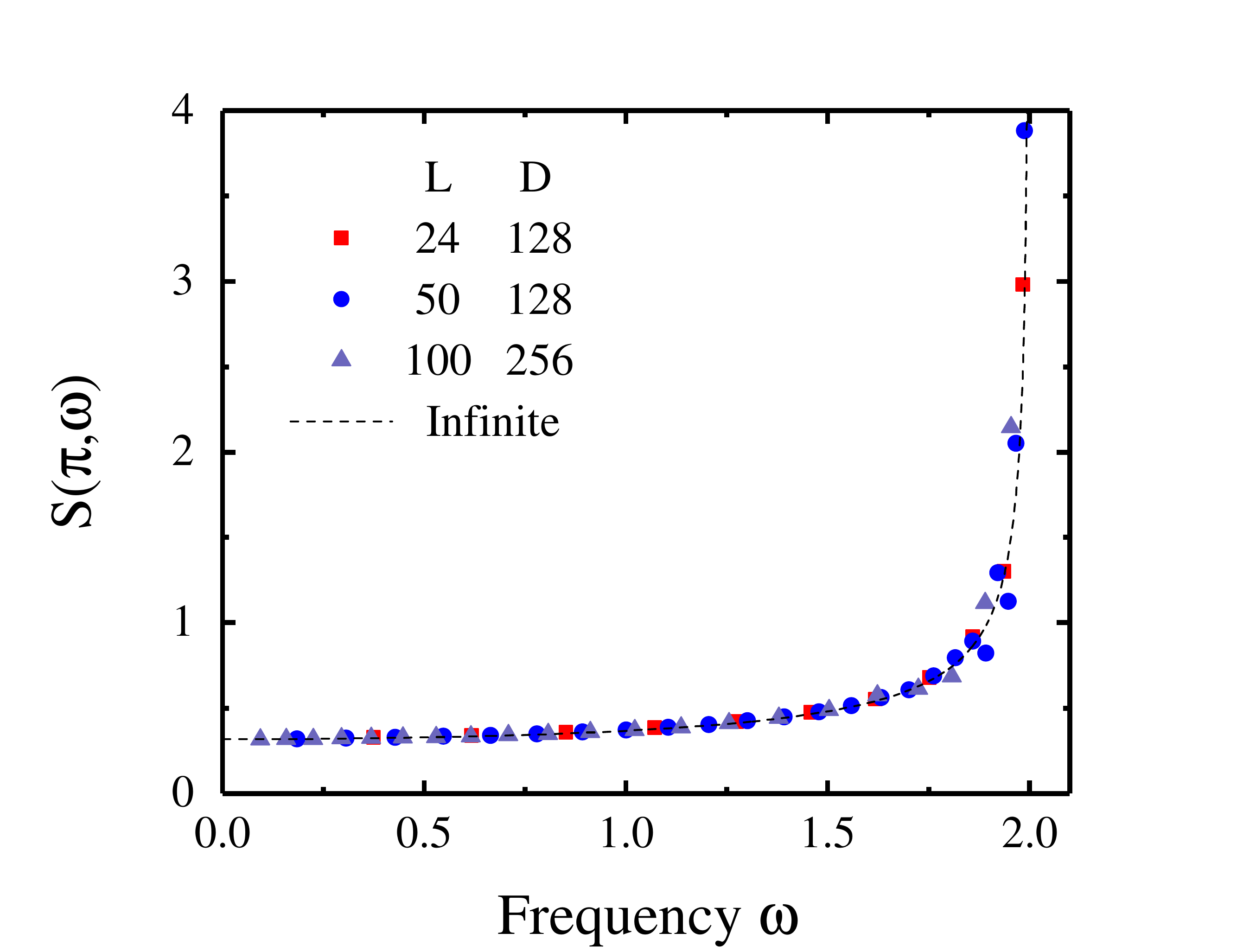}
  \caption{Energy dependence of $S(\pi, \omega)$ obtained using Eq.~(\ref{Eq:SpecFunc}) for the XY model with $L=24$, 50 and 100, corresponding $D=128$, $128$ and $256$. The dash line denotes the exact result in the thermodynamic limit.}
  \label{Fig:XYFinite}
\end{figure}

\begin{figure}[t]
\centering
\includegraphics[width=8.0cm]{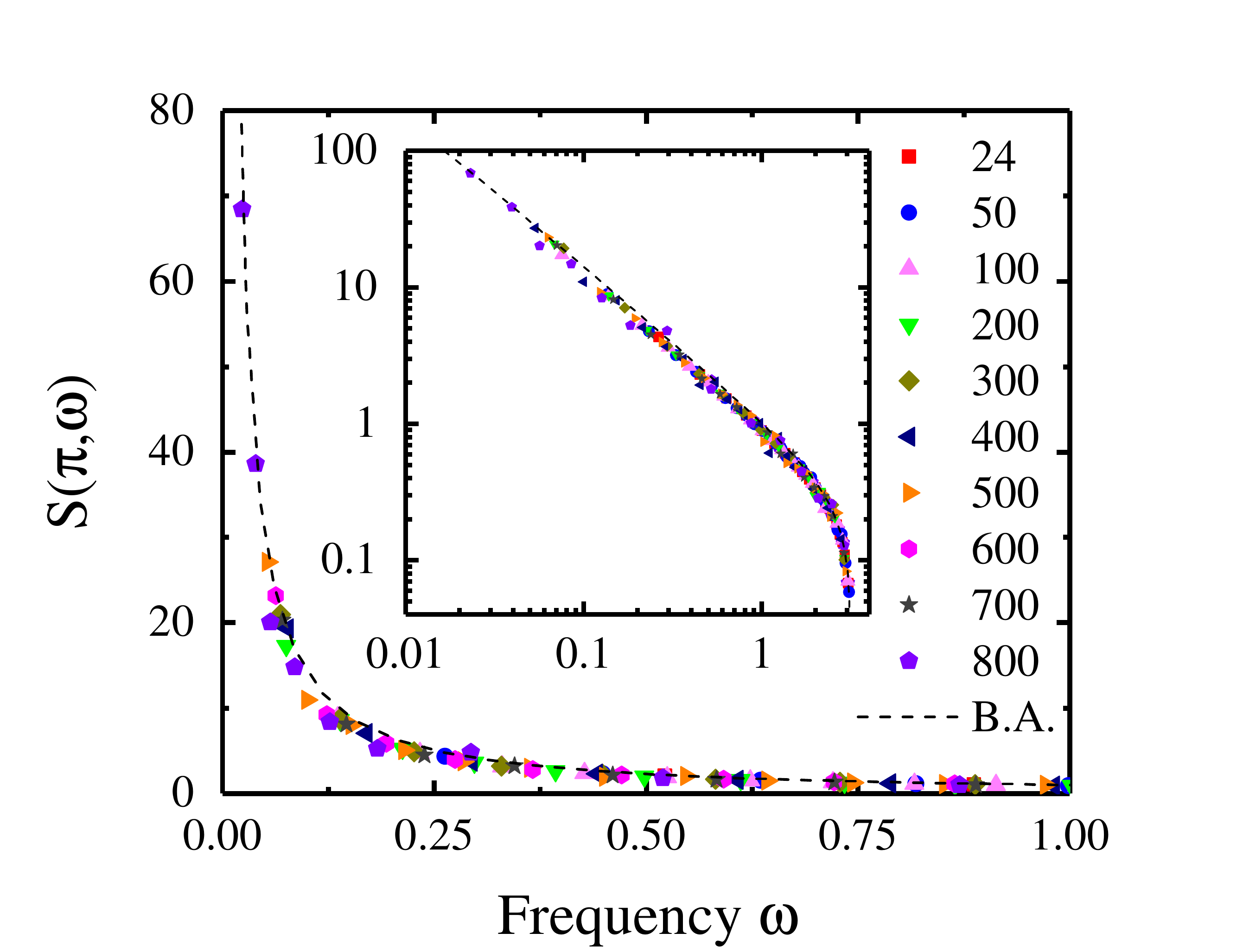}
  \caption{Energy dependence of $S(\pi, \omega)$ obtained using Eq.~(\ref{Eq:SpecFunc}) for the Heisenberg model with the system size $L$, bond dimension $D$ and Chebyshev polynomials $N$ varying as $(24,128,200)$, $(50,128,200)$, $(100,128,100)$, $(200,256,100)$, $(300,256,80)$, $(400,256,60)$, $(500,128,80)$, 
  $(600,128,60)$, $(700,128,50)$ and $(800,64,200)$.
  The inset is the log-log plot of the data.
  The dash line denotes the result of Bethe ansatz (B.A.) in the $L\rightarrow\infty$ limit. }
\label{Fig:HeisFinite}
\end{figure}

We have calculated the dynamical spin-spin correlation functions using Eq.~(\ref{Eq:SpecFunc}) for the XY and Heisenberg models on finite lattice systems.
The results are depicted in Figs.~\ref{Fig:XYFinite} and~\ref{Fig:HeisFinite} for these two models, and compared with the exact results obtained in the limit $L\rightarrow \infty$. For both models, we find that the results we obtain on the finite size lattices using the approach introduced above agree well in the entire energy range with the exact ones obtained in the thermodynamic limit.
Thus our approach provides an accurate and reliable tool to perform the finite size scaling.

\section{Summary} \label{sec:conclusion}

In summary, we propose an orthogonalization scheme to improve the accuracy in the calculation of dynamical spectra functions using the MPS in the Chebyshev expansion.
We first perform a CheMPS calculation to obtain a set of Chebyshev vectors represented by MPS. These recursively generated Chebyshev vectors provide a set of truncated basis states to resolve the spectral function over the entire spectral width uniformly.
However, these vectors do not truly satisfy the recurrence relations of the Chebyshev polynomials due to the approximation used in the CheMPS calculation.
Our approach is to reorthonormalize these vectors to obtain a set of orthonormalized basis states.
The matrix elements of the Hamiltonian are evaluated in this truncated basis space, which defines an effective Hamiltonian.
We then calculate the spectral function using this effective Hamiltonian.
Again the Chebyshev expansion is used to resolve the spectral peaks in a finite lattice system.
Since the effective Hamiltonian is diagonalizable and the Chebyshev polynomials can be determined to an arbitrarily high order, this scheme provides a high-resolution solution to the calculation of dynamic correlation functions.
Using this approach, we have calculated the spectral functions using our method for the 1D XY and Heisenberg models.
By comparison with the results obtained using CheMPS, we find that our method improves greatly both the accuracy and the resolution of dynamical spectra functions.

We can eliminate the broadening effect by taking a discrete representation of the spectral function using directly the eigen-energies of the effective Hamiltonian and the corresponding spectral weight.
The value of spectral function is determined by the ratio between the spectral weight and the corresponding energy width for each eigen-energy.
This can reduce the finite size effect and provide a reliable and accurate approach to scale the finite-lattice results to an infinite lattice.
Our results obtained with this approach, as shown in Figs.~\ref{Fig:XYFinite} and~\ref{Fig:HeisFinite}, agree very well with the exact results obtained in the thermodynamic limit.

Our method works not just in one dimension, it can be generalized to higher dimensions, either using MPS or using other kinds of tensor network states, such as projected entangled pair states (PEPS)~\cite{PEPS} or projected entangled simplex states(PESS)~\cite{PhysRevX.4.011025}.

\section*{Acknowledgments}

This work was supported by the National R\&D Program of China (Grant No. 2017YFA0302901) and the National Natural Science Foundation of China (Grants No. 11190024 and No. 11474331).

\bibliography{HaiDongNov}

\end{document}